\documentclass[pdflatex,sn-mathphys-num,iicol]{sn-jnl}
\usepackage{graphicx}%
\usepackage{multirow}%
\usepackage{amsmath,amssymb,amsfonts}%
\usepackage{amsthm}%
\usepackage{mathrsfs}%
\usepackage[title]{appendix}%
\usepackage{xcolor}%
\usepackage{textcomp}%
\usepackage{manyfoot}%
\usepackage{booktabs}%
\usepackage{algorithm}%
\usepackage{algorithmicx}%
\usepackage{algpseudocode}%
\usepackage{listings}%
\usepackage{orcidlink}
\usepackage[version=4]{mhchem}

\begin{document}

\title[Article Title]{Precision measurements of muonium and muonic helium hyperfine structure at J-PARC}


%\author*[1,2]{\fnm{First} \sur{Author}}\email{iauthor@gmail.com}



%\affil*[1]{\orgdiv{Department}, \orgname{Organization}, \orgaddress{\street{Street}, \city{City}, \postcode{100190}, \state{State}, \country{Country}}}

%\affil[2]{\orgdiv{Department}, \orgname{Organization}, \orgaddress{\street{Street}, \city{City}, \postcode{10587}, \state{State}, \country{Country}}}

%\affil[3]{\orgdiv{Department}, \orgname{Organization}, \orgaddress{\street{Street}, \city{City}, \postcode{610101}, \state{State}, \country{Country}}}

\author*[1,2,3]{\fnm{Patrick} \sur{Strasser}\,\orcidlink{0000-0002-2370-2166}}\email{patrick.strasser(at)kek.jp}

\author[4]{\fnm{Mitsushi} \sur{Abe}\,\orcidlink{0000-0001-5646-7505}}

\author[5]{\fnm{Kanta} \sur{Asai}}

\author[6]{\fnm{Seiso} \sur{Fukumura}}

\author[5]{\fnm{Mahiro} \sur{Fushihara}}

\author[5]{\fnm{Yu} \sur{Goto}}

\author[1,2]{\fnm{Takashi} \sur{Ino}\,\orcidlink{0000-0002-0318-8219}}

\author[7]{\fnm{Ryoto} \sur{Iwai}\,\orcidlink{0009-0002-3645-5683}}

\author[1,2,3]{\fnm{Sohtaro} \sur{Kanda}\,\orcidlink{0000-0002-9080-4154}}

\author[5]{\fnm{Shiori} \sur{Kawamura}\,\orcidlink{0009-0006-5501-0101}}

\author[5,8,1]{\fnm{Masaaki} \sur{Kitaguchi}\,\orcidlink{0000-0001-7637-6467}}

\author[1,2]{\fnm{Shoichiro} \sur{Nishimura}\,\orcidlink{0000-0002-5581-4090}}

\author[9,10]{\fnm{Takayuki} \sur{Oku}\,\orcidlink{0000-0001-8042-6113}}

\author[5,9]{\fnm{Takuya} \sur{Okudaira}\,\orcidlink{0000-0002-5150-9275}}

\author[11]{\fnm{Adam} \sur{Powell}\,\orcidlink{0000-0003-2475-6067}}

\author[4]{\fnm{Ken-ichi} \sur{Sasaki}\,\orcidlink{0000-0002-6354-3876}}

\author[5,1]{\fnm{Hirohiko M.} \sur{Shimizu}\,\orcidlink{0000-0002-8767-6542}}

\author[1,2,3]{\fnm{Koichiro} \sur{Shimomura}\,\orcidlink{0000-0002-9501-6149}}

\author[5]{\fnm{Hiroki} \sur{Tada}\,\orcidlink{0000-0002-3722-7095}}

\author[12]{\fnm{Hiroyuki A.} \sur{Torii}}

\author[13]{\fnm{Takashi} \sur{Yamanaka}\,\orcidlink{0009-0008-4654-9097}}

\author[1,2,3]{\fnm{Takayuki} \sur{Yamazaki}\,\orcidlink{0000-0003-4771-8821}}

\author[ ]{(MuSEUM~Collaboration)}

%%Orcid ID:
% Patrick Strasser    	{0000-0002-2370-2166}
% Takashi Ino         	{0000-0002-0318-8219}
% Ryoto Iwai          	{0009-0002-3645-5683}
% Sohtaro Kanda       	{0000-0002-9080-4154}
% Shiori Kawamura     	{0009-0006-5501-0101}
% Masaaki Kitaguchi   	{0000-0001-7637-6467}
% Shoichiro Nishimura 	{0000-0002-5581-4090}
% Takayuki Oku        	{0000-0001-8042-6113}
% Takuya Okudaira     	{0000-0002-5150-9275}
% Adam Powell:        	{0000-0003-2475-6067}
% Ken-ichi Sasaki      	{0000-0002-6354-3876}
% Hirohiko M. Shimizu 	{0000-0002-8767-6542}
% Koichiro Shimomura  	{0000-0002-9501-6149}
% Hiroki Tada         	{0000-0002-3722-7095}
% Mitsushi Abe        	{0000-0001-5646-7505}
% Takashi Yamanaka:   	{0009-0008-4654-9097}
% Takayuki Yamazaki   	{0000-0003-4771-8821}

\affil[1]{\orgdiv{Institute of Materials Structure Science (IMSS)},
          \orgname{High Energy Accelerator Research Organization (KEK)},
          \orgaddress{
          \street{1-1 Oho},
          \city{Tsukuba},
          \state{Ibaraki}
          \postcode{305-0801},
          \country{Japan}}}

\affil[2]{\orgdiv{Materials and Life Science Division},
          \orgname{J-PARC Center},
          \orgaddress{
          \street{2-4 Shirakata, Tokai-mura},
          \city{Naka-gun},
          \state{Ibaraki}
          \postcode{ 319-1195},
          \country{Japan}}}

\affil[3]{\orgdiv{Materials Structure Science Program},
          \orgname{Graduate Institute for Advanced Studies, SOKENDAI},
          \orgaddress{
          \street{1-1 Oho},
          \city{Tsukuba},
          \state{Ibaraki}
          \postcode{305-0801},
          \country{Japan}}}

\affil[4]{\orgdiv{Cryogenics Science Center},
          \orgname{High Energy Accelerator Research Organization (KEK)},
          \orgaddress{
          \street{1-1 Oho},
          \city{Tsukuba},
          \state{Ibaraki}
          \postcode{305-0801},
          \country{Japan}}}

\affil[5]{\orgdiv{Graduate School of Science},
          \orgname{Nagoya University},
          \orgaddress{
          \street{Furo-cho, Chikusa-ku},
          \city{Nagoya},
          \state{Aichi}
          \postcode{464-8601},
          \country{Japan}}}

\affil[6]{\orgdiv{Department of Physics},
          \orgname{Niigata University},
          \orgaddress{\street{8050 Ikarashi 2-no-cho, Nishi-ku},
          \city{Niigata-shi},
          \state{Niigata}
          \postcode{950-2181},
          \country{Japan}}}

\affil[7]{\orgdiv{Facility for Rare Isotope Beams},
          \orgname{Michigan State University},
          \orgaddress{
          \street{640 South Shaw Lane},
          \city{East Lansing},
          \state{Michigan}
          \postcode{48824},
          \country{USA}}}

\affil[8]{\orgdiv{Kobayashi-Maskawa Institute},
          \orgname{Nagoya University},
          \orgaddress{
          \street{Furo-cho, Chikusa-ku},
          \city{Nagoya},
          \state{Aichi}
          \postcode{464-8601},
          \country{Japan}}}

\affil[9]{\orgdiv{Advanced Science Research Center},
          \orgname{Japan Atomic Energy Agency (JAEA)},
          \orgaddress{
          \street{2-4 Shirakata, Tokai-mura,},
          \city{Naka-gun},
          \state{Ibaraki}
          \postcode{319-1195},
          \country{Japan}}}

\affil[10]{\orgdiv{Graduate School of Science and Engineering},
          \orgname{Ibaraki University},
          \orgaddress{
          \street{2-1-1 Bunkyo},
          \city{Mito},
          \state{Ibaraki}
          \postcode{310-8512},
          \country{Japan}}}

\affil[11]{\orgdiv{Department of Physics and Astronomy},
          \orgname{University of Calgary},
          \orgaddress{
          \street{2500 University Drive NW},
          \city{Calgary},
          \state{Alberta}
          \postcode{T2N 1N4},
          \country{Canada}}}

\affil[12]{\orgdiv{School of Science},
          \orgname{The University of Tokyo},
          \orgaddress{
          \street{7-3-1 Hongo},
          \city{Bunkyo-ku},
          \state{Tokyo}
          \postcode{113-0033},
          \country{Japan}}}

\affil[13]{\orgdiv{Faculty of Arts and Science},
          \orgname{Kyushu University},
          \orgaddress{
          \street{744 Motooka},
          \city{Nishi},
          \state{Fukuoka}
          \postcode{819-0395},
          \country{Japan}}}

\abstract{
At the J-PARC Muon Science Facility (MUSE), the MuSEUM collaboration is now performing new precision measurements of the ground state hyperfine structure (HFS) of both muonium and muonic helium atoms. 
High-precision measurements of the muonium ground-state HFS are recognized as one of the most sensitive tools for testing bound-state quantum electrodynamics theory to precisely probe the standard model and determine fundamental constants of the positive muon magnetic moment and mass. 
The same technique can also be employed to measure muonic helium HFS, obtain the negative muon magnetic moment and mass, and test and improve the theory of the three-body atomic system.
Measurements at zero magnetic field have already yielded more accurate results than previous experiments for both muonium and muonic helium atoms.
High-field measurements are now ready to start collecting data using the world's most intense pulsed muon beam at the MUSE \mbox{H-line}. 
We aim to improve the precision of previous measurements ten times for muonium and a hundred times or more for muonic helium.
We review all the key developments for these new measurements, focusing on the high-field experiment, and report the latest results and prospects.
}




\maketitle

%--------------------------------------------------------------------------------
\section{Introduction}\label{introduction}

Particle physics aims to reveal the ultimate state of matter at the microscopic level and the mechanisms of its interactions. The standard model of elementary particles explains almost all of the myriad physics experiments on Earth.
However, the ultimate theory has not yet been reached, as it does not include subatomic particles corresponding to dark matter, whose existence has now been shown by cosmic observations~\cite{HyeongHan2024}.
The existence of new physics should be manifested as faint signs, which will become observable if the measurement precision and sensitivity of experiments are improved.

The muon, a second-generation charged lepton, is the most suitable particle for detecting signs of new physics because of its moderate mass and lifetime.
Modern precision experiments of charged particles interacting with electromagnetic fields have extremely high accuracy in frequency measurements, which is well suited for muons.
In recent years, several physical phenomena involving muons have been observed that could be explained.

The first is the large deviation (7$\sigma$) of the proton radius values obtained from laser spectroscopy of muonic hydrogen atoms (a bound state of a negative muon and a proton) from those obtained from proton and electron inelastic scattering experiments and hydrogen atom spectroscopy (known as the proton radius puzzle)~\cite{Antognoni2013}. Currently, this situation seems to have been resolved with new or re-analyzed measurements that agree with the muon measurement~\cite{Workman2022}.

The second is the large discrepancy (5$\sigma$) between theory and measurement in the deviation of the muon $g$ factor from 2 (known as muon $g-2$). 
This difference did not disappear even after the first new measurement in 20 years~\cite{Aguillard2023}, making it one of the most interesting physics topics in recent years.
New developments, especially in lattice QCD calculations~\cite{Bazavov2024arXiv} and new measurements of the pion form factor~\cite{Ignatov2024}, seem to have closed this gap, but it is still too early to make a statement.
The standard model theory prediction is still a work in progress.
The important lesson is that muon experiments have pointed to the problems in both cases: missing pieces or sign new physics.

We report on new precision microwave spectroscopy measurements of two exotic hydrogenlike atoms, muonium (Mu) and muonic helium ($\mu$He), which are composed of positive and negative muons, respectively. 
Muonium will be used to validate the quantum electromagnetic dynamics (QED) theory, the core of the standard model. 
The accuracy of the mass measurements for both positive and negative muons will be improved, ultimately allowing an experimental verification of the $CPT$ invariance with second-generation leptons.

%--------------------------------------------------------------------------------
\section{Muonium and muonic helium atoms}

\begin{figure}[ht]
  \centering
  \includegraphics[width=76mm]{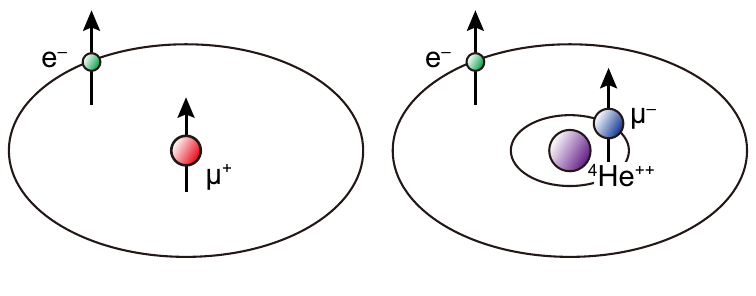}
  \caption{Comparison between a muonium atom (left) and a muonic helium atom (right).}
  \label{Fig1}
\end{figure}

The MuSEUM\footnote{MuSEUM stands for ``Muonium Spectroscopy Experiment Using Microwave''.} collaboration is now performing new precision measurements of the hyperfine structure (HFS) of both muonium and muonic helium atoms~\cite{Shimomura2011, Strasser2018} at the Muon Science Facility (MUSE)~\cite{Higemoto2017} of the Japan Proton Accelerator Research Complex (J-PARC). 
Muonium (a bound state of a positive muon and an electron) and muonic helium (a helium atom with one of its electrons replaced by a negative muon) are both hydrogen-like atoms (see Fig.~\ref{Fig1}). Their respective ground-state HFS results from the interaction of the muon and electron magnetic moments, which are very similar but inverted because of the different signs of their respective muon magnetic moments. 
The same technique can be employed to measure both muonium and muonic helium HFS. Although these exotic atoms share many similarities in terms of experimental apparatus, they are quite different physical systems.

%Fig1---------------------------------------------------------
%-------------------------------------------------------------

%---------------------------------------
\subsection{Muonium HFS}

High-precision measurement of the muonium ground state hyperfine structure $\Delta\nu_{\text{HFS}}^{\text{Mu}}$ is regarded as one of the most sensitive tools for testing QED theory and determining the fundamental constants of the positive muon magnetic moment $\mu_{\mu^{+}}$ and mass $m_{\mu^{+}}$ ($\propto 1/\mu_{\mu^{+}}$).
Muonium is the most suitable system for testing QED because both theoretical and experimental values can be precisely determined. 
Indeed, its lifetime of 2.2 $\mu$s is long enough in comparison to positronium~\cite{ Ishida2014, Baker2014} to measure it precisely, and being a purely leptonic system (neither of its constituents has an internal structure) as compared to hydrogen~\cite{Ramsey1990, Eides2007} it can be accurately calculated. 
%\cite{Ritter1984, Mills1983, Ishida2014, Baker2014},~\cite{Essen1973, Eides2007}

Currently the muonium hyperfine splitting $\Delta\nu_{\text{HFS}}^{\text{Mu}}$ = 4~463~302~765(53)~Hz is determined with an experimental precision of 12~ppb, and the ratio of the muon and proton magnetic moments $\mu_{\mu^{+}}/\mu_{p}$ at 120~ppb from the previous high-field experiment at LAMPF~\cite{Liu1999} mostly dominated by statistical errors.
By combining the world's most intense pulsed muon beam at the MUSE \mbox{H-line}~\cite{Kawamura2014, Yamazaki2023} with a new analysis method ``Rabi oscillation spectroscopy''~\cite{Nishimura2021} (see Sect.~\ref{analysis}), we aim at measuring $\Delta\nu_{\text{HFS}}^{\text{Mu}}$ and $\mu_{\mu^{+}}/\mu_{p}$ with an order of magnitude higher precision.

The largest uncertainty in the muonium HFS theoretical value~\cite{Eides2019} comes from the experimental determination of the muon magnetic moment and its mass.
A ten times improvement will allow for the first time to experimentally probe the hadronic vacuum-polarization ($\sim$~230~Hz) and electroweak ($\sim$~65~Hz) contributions. 
Hadronic vacuum polarization contributes to the largest theoretical uncertainty for the muon $g-2$~\cite{Shintani2019} and also plays a part in the refinement of the proton Zemach radius, needed for a quantitative understanding of the Lamb shift (proton radius puzzle) of muonic hydrogen atoms~\cite{Antognoni2013, Pohl2016}.
The goal is to compare the theoretical prediction for the muonium HFS with the new experimental results in search of new physics. A discrepancy between theory and experiment could be interpreted as new physics. 
Thus, the theoretical value of Eides~\cite{Eides2019} is preferred to CODATA adjustment value~\cite{Mohr2024arXiv} to take into account the proper magnitude of the error bars of the theoretical prediction for muonium HFS (see also the discussion in~\cite{Karshenboim2021}). 
Underestimating the error bars could result in a new physics discovery being misreported.

Furthermore, the magnetic moment ratio between muon and proton $\mu_{\mu^{+}}/\mu_{p}$ is needed in the determination of the muon anomalous magnetic moment in the muon $g-2$ experiment~\cite{Bennett2006, Aguillard2023}. 
This is because proton nuclear magnetic resonance (NMR) is used for magnetic field measurement in both MuSEUM and $g-2$ experiments.
Precision microwave spectroscopy of muonium can also contribute to numerous new physics, such as testing $CPT$ and Lorentz invariance incorporated in extensions to the standard model~\cite{Hughes2001, Kostelecky2015}, 
%disentangling the proton radius puzzle~\cite{Antognoni2013, Pohl2016}, 
searching for exotic particle~\cite{Karshenboim2010, Karshenboim2011}, probing for long-range neutrino-mediated forces~\cite{Stadnik2018}, and searching for ultralight scalar dark matter~\cite{Stadnik2023}.

%---------------------------------------
\subsection{Muonic helium HFS}

A muonic helium atom is formed when a negative muon is stopped in helium gas and captured by a helium atom. 
The Bohr radius of the bound $\mu^{-}$ in helium is $\sim$400~times smaller than that of a hydrogen atom. The muon is so closely bound to the helium nucleus that it nearly completely screens one proton charge, producing a ``pseudonucleus'' [$\mu^{-}$$^{4}$He$^{++}$]$^{+}$ with a positive effective charge and a magnetic moment nearly equal to that of a negative muon $\mu_{\mu^{-}}$.
Thus, it can be regarded as a heavy hydrogen isotope, similar to muonium, and forms with it the longest isotopic chain (mass ratio of 36).
As mentioned earlier, the muonic helium ground-state hyperfine structure $\Delta\nu_{\text{HFS}}^{\mu\text{He}}$, which results from the interaction of the negative muon magnetic moment and the remaining electron, is very similar to that of muonium but inverted. 
The same technique as with muonium can be used to precisely measure $\Delta\nu_{\text{HFS}}^{\mu\text{He}}$, which is a sensitive tool to test three-body atomic systems and bound-state QED theories. This also allows us to determine the fundamental constants of the negative muon magnetic moment $\mu_{\mu^{-}}$ and mass $m_{\mu^{-}}$ ($\propto 1/\mu_{\mu^{-}}$).
Muonic helium HFS has only been measured twice in the 1980s, directly at zero magnetic field~\cite{Orth1980} and indirectly at high field~\cite{Gardner1982}. 
Currently $\Delta\nu_{\text{HFS}}^{\mu\text{He}}$ is determined with an experimental precision of 6.5~ppm, and $\mu_{\mu^{-}}/\mu_{p}$ at 47~ppm, mainly dominated by statistical errors~\cite{Gardner1982}.

What makes muonic helium HFS measurements more challenging is that the initial $\mu^{-}$ polarization ($\sim$100\%) is strongly reduced down to 2--5\%~\cite{Souder1980, Orth1984} during the muon cascade process in He due to Auger transition and collisional Stark mixing.
This value should be compared to the case of muonium (50\%), thus making it more challenging to measure the HFS interval as we need to measure a hundred times longer to get similar statistics with muonic helium.
Also, when a He atom captures a $\mu^{-}$ ejecting both electrons via Auger transitions in the process, and a ($\mu$$^{4}$He)$^{+}$ ion in its ground 1$s$ state is formed, it cannot capture an electron from neighboring He atoms because its electron binding energy is similar to that in hydrogen (13.6~eV).
Thus, to form a neutral $\mu$He atom, the prerequisite to measuring HFS, a collision with a foreign gas atom acting as an electron donor is necessary.
Here, CH$_{4}$ was preferred to Xe used previously~\cite{Orth1980, Gardner1982} because of its reduced total charge ($Z=10$) making it less likely to capture muons (Fermi-Teller $Z$~law), a similar ionization potential (12.5~eV), and is also believed to give a larger residual polarization compared to Xe~\cite{Arseneau2016}, making it advantageous for the measurements.

The ground state HFS of muonic helium is very similar to that of muonium, but in reality, it is complicated because three-body interactions have to be considered, thus limiting the theoretical approach. 
$\Delta\nu_{\text{HFS}}^{\mu\text{He}}$ is $\sim$~1.7~kHz (380~ppm) larger than $\Delta\nu_{\text{HFS}}^{Mu}$ due to these interactions. 
Muonic helium HFS is the only available experimental data for three-body muonic atoms. Unfortunately, the latest theoretical value~\cite{Aznabayev2018} is still 30~times less precise.
However, there is hope that new theoretical calculations developed for HFS in $^3$He~\cite{Patkos2023} could be applied to muonic helium HFS. This would improve the current theory to the same level as the present experimental accuracy and give the first opportunity to test QED effects in three-body muonic atoms. 
We are confident that these new measurements will motivate QED theorists.

Test of the $CPT$ theorem with muons is now limited by the accuracy of the negative muon mass to a level of 3~ppm~\cite{Fei1994}, which should be compared to that of the positive muon mass of 120~ppb~\cite{Liu1999}.
The positive muon mass will soon be improved ten times by the muonium HFS measurements. 
Also, several other experiments are now in progress by the muonium 1$s$--2$s$ spectroscopy experiment at J-PARC~\cite{Uetake2019}, and MuMASS at PSI~\cite{Crivelli2018}. 
A more precise $\mu_{\mu^{-}}$/$\mu_{p}$ would also be welcome to determine the negative muon magnetic moment anomaly $a_{\mu^{-}}$ in the muon $g-2$ experiment at Brookhaven National Laboratory. 
The more accurate $\mu_{\mu^{+}}$/$\mu_{p}$~\cite{Liu1999} is currently used for both $a_{\mu^{+}}$ and $a_{\mu^{-}}$ to test the standard model's predictions and $CPT$ theorem~\cite{Bennett2006}. 

%----------

We aim at improving previous measurements nearly a hundred times by combining the high-intensity pulsed negative muon beam at MUSE \mbox{H-line}, the increased detection efficiency of decay electrons being more focused on the upstream and downstream detectors by the high-magnetic field, and utilizing the Rabi-oscillation spectroscopy method~\cite{Nishimura2021}, reaching a precision for $\Delta\nu_{\text{HFS}}^{\mu\text{He}}$ below 100~ppb after 100 days.
This will also allow to determine $\mu_{\mu^{-}}$ and $m_{\mu^{-}}$ below 1~ppm to test $CPT$ invariance by comparing with positive muons. 
A more precise determination of the muonic helium atom HFS will be beneficial to test and improve the theory of the three-body atomic system.

Furthermore, we are now investigating a new experimental approach to recover the polarization lost during the muon cascade by repolarizing $\mu$He atoms using a spin-exchange optical pumping technique \cite{Barton1993, Fukumura2024}, which could further improve the measurement precision by nearly 1 order of magnitude, reaching ultimately $\mathcal{O}$(10~ppb) for $\Delta\nu_{\text{HFS}}^{\mu\text{He}}$ and $\mathcal{O}$(100~ppb) for $\mu_{\mu^{-}}$.

%--------------------------------------------------------------------------------
\section{Experimental method}

\begin{figure}[t]
  \centering
  \includegraphics[width=76mm]{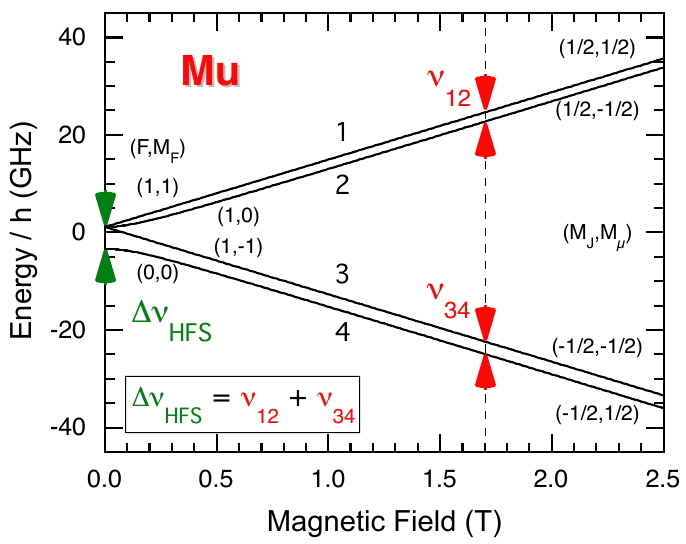}
  \includegraphics[width=76mm]{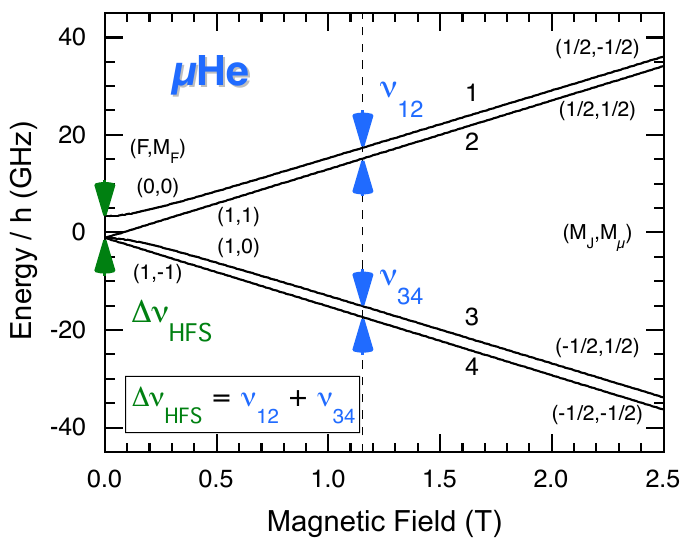}
  \caption{Breit-Rabi energy level diagram for muonium atom (top) and muonic helium atom (bottom) under high magnetic field.}
  \label{Fig2}
\end{figure}

The muonium ground-state hyperfine splitting $\Delta\nu_{\text{HFS}}^{\text{Mu}}$ is measured by a microwave magnetic resonance technique. This transition can be measured directly at zero field and indirectly following the approach of the last experiment~\cite{Liu1999} in a static magnetic field. 
The Breit-Rabi energy level diagram for the muonium atom (Mu) is shown in Fig.~\ref{Fig2}(top). 
The static magnetic field due to the Zeeman effect splits the ground state of muonium into four different substates, and the two spin-flip resonance frequencies $\nu_{12}$ and $\nu_{34}$ are measured. 
Since two resonance frequencies are measured, two physical constants can be determined. 
The sum of these two transition frequencies is constant, independent of the applied static field, and equal to the ground state HFS at zero field, i.e., $\Delta\nu_{\text{HFS}}^{\text{Mu}} = \nu_{12}$ + $\nu_{34}$. 
The difference between these two transition frequencies is directly related to the ratio of the muon and proton magnetic moments, i.e., $\mu_{\mu^{+}}/\mu_{p} \propto \nu_{34} - \nu_{12}$. 
Thus, the sum of the measured frequencies gives the QED test, while the difference is used to determine the values of the positive muon magnetic moment and mass.

%Fig2---------------------------------------------------------
%-------------------------------------------------------------

The same technique can be used to measure the ground state hyperfine structure of muonic helium $\Delta\nu_{\text{HFS}}^{\mu\text{He}}$. 
The Breit-Rabi energy level diagram for the muonic helium atom is shown in Fig.~\ref{Fig2}(bottom). 
The diagram is inverted because of the different signs of the respective muon magnetic moments. 
In this case, the sum of the measured frequencies gives a QED test of the three-body atomic system, and the difference is used to determine the values of the negative muon magnetic moment and mass.

%---------------------------------------
\subsection{Measurement procedure}

\begin{figure}[t]
  \centering
  \includegraphics[width=76mm]{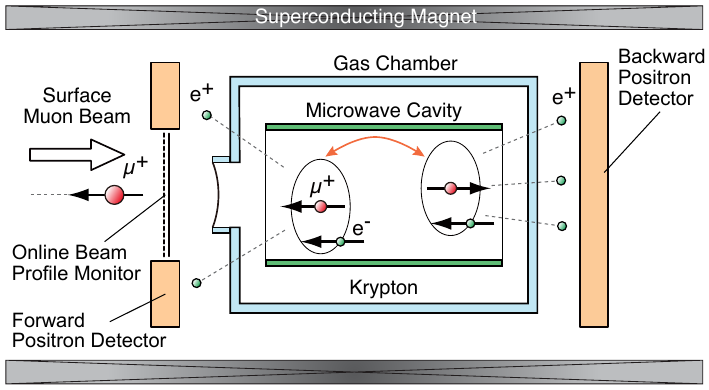}
  \caption{Schematic view of the experimental arrangement inserted in a superconducting magnet to measure muonium HFS at high field.}
  \label{Fig3}
\end{figure}

The schematic view of the experimental apparatus inserted in a large superconducting solenoid for high-field measurements is shown in Fig.~\ref{Fig3}.
%with the applied static field parallel to the muon momentum direction.
The experimental procedure is (1) muonium formation, (2) microwave spin flip, and (3) positron asymmetry measurement. 
High-intensity surface ($\mu^+$) muons, 100\% backward polarized with respect to the muon momentum direction, are injected into a microwave cavity located inside a gas chamber containing highly pure krypton gas. 
The profile of the incident muon beam is measured online by a non-destructive beam profile monitor located in front of the entrance window of the gas chamber. 
The $\mu^+$ are stopped in the Kr gas volume and form polarized muonium atom through the charge exchange reaction $\mu^+$ + Kr $\rightarrow$ Mu + Kr$^+$. 
The muon spin can be flipped by applying a microwave magnetic field in the microwave cavity perpendicular to the muon direction. 
The positrons (e$^+$) from $\mu^+$ decay are emitted preferentially in the direction of the muon spin. 
At the resonance, the microwave field induces the muon spin flip, changing the angular distribution of the emitted positrons from primarily backward to forward direction. 
Positrons are then detected with segmented scintillation detectors placed downstream and upstream of the gas chamber. 
Muonium HFS measurements are performed by scanning the microwave frequency and measuring the decay $e^{+}$ asymmetry with and without microwave (N\textsubscript{ON}/N\textsubscript{OFF} $-$ 1) to determine the resonance frequency, i.e., $\Delta\nu_{\text{HFS}}^{\text{Mu}}$ at zero field and $\nu_{12}$ and $\nu_{34}$ at high field, respectively.

%Fig3---------------------------------------------------------
%-------------------------------------------------------------

The same experimental apparatus is used to measure the muonic helium HFS. 
The main differences are (1) the injection of polarized negative muons ($\mu^{-}$), (2) the gas chamber is filled with pressurized helium gas containing 2\% CH$_{4}$ admixture as an electron donor to form neutral muonic helium atoms efficiently, and (3) electrons (e$^-$) from $\mu^-$ decay are preferentially emitted in the direction opposite to the negative muon spin.

%---------------------------------------
\subsection{Measurement strategy}

\begin{figure}[b]
  \centering
  \includegraphics[width=76mm]{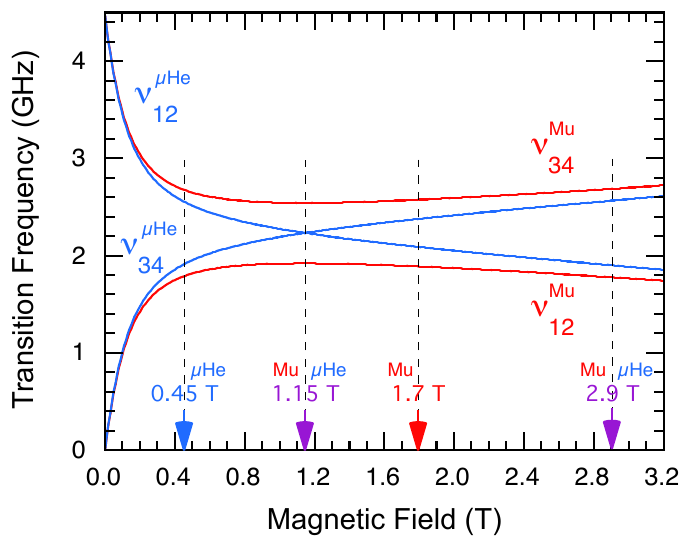}
  \caption{HFS transition frequencies for muonium (Mu) and muonic helium ($\mu$He) atoms as a function of the applied static magnetic field. The arrows show the measurements planned at different magnetic fields.}
  \label{Fig4}
\end{figure}

Performing measurements at different magnetic fields can potentially improve the experimental results, for instance, the uncertainty of $\mu_{\mu}/\mu_{p}$ can be reduced, and the systematic uncertainty due to the magnetic field uncertainty can be cross-checked.
Figure~\ref{Fig4} shows the dependence of the resonance frequencies $\nu_{12}$ and $\nu_{34}$ as a function of the applied static magnetic field for both muonium and muonic helium atoms.

%Fig4---------------------------------------------------------
%-------------------------------------------------------------

The muonium resonance frequencies $\nu_{12}^{\text{Mu}}$ and $\nu_{34}^{\text{Mu}}$ will first be measured in a magnetic field of 1.7~T following the approach of the last experiment~\cite{Liu1999}. 
Then, the magnetic field uncertainty can be reduced by measuring at the maximum field strength produced by our MRI magnet (2.9~T) since the magnetic probe accuracy is independent of the applied field. 
Finally, we plan to measure at a field of $\approx$~1.15 T, where the effect of the fluctuating magnetic field is minimal because the derivative of $\nu_{12}^{\text{Mu}}$ and $\nu_{34}^{\text{Mu}}$ becomes zero. 
Combining these three measurements, which will require different microwave cavities (see Sect.~\ref{microwave}), will put a stringent test on the internal consistency of the experiment's systematic uncertainty. 

In the case of muonic helium atoms, both transition frequencies $\nu_{12}^{\mu\text{He}}$ and $\nu_{34}^{\mu\text{He}}$ are equal at a magnetic field of $\approx$~1.15 T and can be driven simultaneously using a cylindrical cavity with a single TM110 mode, similar to the previous experiment~\cite{Gardner1982}.
Moreover, $\nu_{12}^{\mu\text{He}}$ and $\nu_{34}^{\mu\text{He}}$ at a magnetic field of 0.45~T and 2.9~T are nearly identical to those of muonium at 1.7~T, thus the same microwave cavity can be used for these measurements.

%--------------------------------------------------------------------------------
\section{Experimental apparatus}

The high-field measurements are nearly the same as at zero field. 
What is required, in addition, is a superconducting magnet to generate a uniform high-magnetic field, a high-precision magnetic probe, and a cavity that resonates with two different microwave frequencies. 
A blind analysis method is also needed to eliminate human bias to achieve the desired accuracy.

%---------------------------------------
\subsection{Uniform magnetic field and magnetic probes}

A superconducting solenoid recycled from an old MRI magnet with a 680~mm effective bore diameter will be used to provide the high-magnetic field homogeneity required for the experiment.
Our choice of a longer cavity to allow measurement at lower Kr gas density imposes strict requirements on the magnetic field. 
We aim at a field homogeneity of less than 0.2~ppm with absolute calibration at 10~ppb levels in a spheroidal volume of $\phi$200~mm $\times$ 300~mm (muon stopping region). 

It should be noted that the systematic uncertainty on the magnetic field was the second largest uncertainty in the previous experiment at LAMPF~\cite{Liu1999}. The muonium HFS frequency $\Delta\nu_{\text{HFS}}$ ($= \nu_{12} + \nu_{34}$) is in principle independent of the magnetic field. However, $\mu_{\mu}$/$\mu_{p}$ ($\propto \nu_{34} - \nu_{12}$) is dependent on the magnetic field, and its inhomogeneity relates largely to the systematic uncertainty. Since both $\nu_{12}$ and $\nu_{34}$ are measured separately, a precise calibration is still required in both cases. 

The fine adjustment of the magnetic field homogeneity was already established using a shimming method by the insertion of thin iron or nickel plates. The shim plate positioning is calculated using a singular value decomposition method, an iterative process of field correction and precise field measurement. 
A simulation confirmed that the final target uniformity could be achieved by using magnetic putty~\cite{Sasaki2022}.
We have also confirmed the long-term stability at 3~ppb/hour over 10~days, with a helium evaporation rate of 3 L/day. After two coarse and three fine iterative shimming processes, a field inhomogeneity of 0.2~ppm peak-to-peak in the muon-stopping region spheroid was achieved in the H1 area, fulfilling our design value.

Three kinds of magnetic probes are used: (1) a standard probe for absolute calibration, (2) a fixed probe for monitoring the magnetic field during the measurement, and (3) a magnetic field camera to map the magnetic field three-dimensionally.

\begin{figure}[t]
  \centering
  \includegraphics[width=76mm]{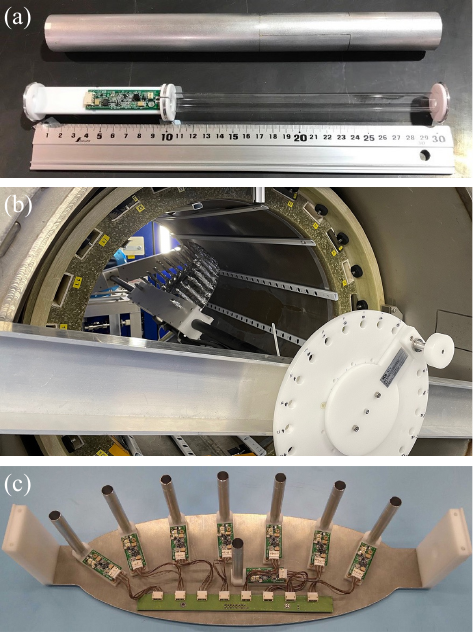}
  \caption{(a) Standard CW-NMR probe, (b) magnetic field camera installed in the MRI magnet, and (c) small magnetic field camera for mapping inside the microwave cavity.}
  \label{Fig5}
\end{figure}

A continuous wave nuclear magnetic resonance (CW-NMR) system with an RF pickup coil was developed to precisely measure the magnetic field for both muon $g-2$/EDM and MuSEUM experiments at J-PARC~\cite{Sasaki2016}. The development of the CW-NMR probe is reported in detail in~\cite{Tanaka2018}.
Figure~\ref{Fig5}(a) shows the developed standard probe, which recently achieved a magnetic field measurement accuracy of 15~ppb. 

A precise magnetic field camera using a set of 24 water NMR probes was developed to scan and measure the magnetic field map of the muon-stopping spheroid at high speed (24~probes $\times$ 24~positions)~\cite{Tada2022}. Figure~\ref{Fig5}(b) shows the magnetic field camera installed in the MRI magnet during the shimming procedure. A smaller field camera was also developed to measure the field map inside the microwave cavity (Fig.~\ref{Fig5}(c)).

%Fig5---------------------------------------------------------
%-------------------------------------------------------------

%---------------------------------------
\subsection{Microwave system}\label{microwave}

\begin{figure}[t]
  \centering
  \includegraphics[width=76mm]{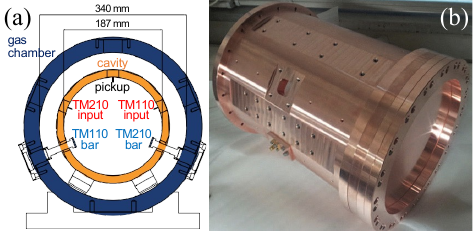}
  \caption{(a) Cross-sectional view and (b) picture of the microwave cavity for high-field measurements.}
  \label{Fig6}
\end{figure}

Muonium transition frequencies in a 1.7~T magnetic field are approximately $\nu_{12} \approx$ 1.897~GHz and $\nu_{34} \approx$ 2.566~GHz.
A cylindrical microwave cavity was already designed and constructed for the high-field measurement to have two frequency modes, TM110 and TM210, to match the resonance frequencies $\nu_{12}$ and $\nu_{34}$~\cite{TanakaThesis2015, Tanaka2021}.
The frequency ratio of TM110 and TM210 modes yields a constant value for a cylindrical cavity, independently of the cavity radius. Thus, a cylindrical cavity can only be used at 1.7 T (magic field), where the resonance frequency relation $f_{110}/f_{210} = \nu_{12}/\nu_{34}$ holds.
As shown in Fig.~\ref{Fig6}(a), there are two input ports, one for each frequency, each equipped with a piezo actuator directly attached to the gas chamber for tuning by displacing a conductive tuning bar, covering the full sweep range of the resonance line shape.
The microwave cavity shown in Fig.~\ref{Fig6}(b) is made of oxygen-free copper with windows 25-$\mu$m thick to let muons in and positrons out. 
The cavity with an 187~mm inner diameter is 304~mm long as compared to 160~mm for the LAMPF experiment to allow measurement at lower Kr gas density, thus reducing the systematic uncertainty on the gas density extrapolation to zero pressure. 
The stored microwave power can be monitored through a pickup antenna with a power meter.
Since resonance lines are distorted by fluctuating microwave power, the signal from the pickup antenna in the cavity will be used to monitor and regulate the power to achieve a stability of 0.01\%~\cite{Tanaka2021}. 

To achieve even higher precision, we also aim to measure muonium HFS at a 2.9~T magnetic field.
The transition frequencies are different from that at 1.7~T, and a cylindrical cavity can no longer be used. Therefore, a dual-mode rectangular microwave cavity with more degrees of freedom in choosing the two-mode frequencies was developed.
A prototype for the measurement at 2.9~T field was designed and constructed~\cite{Iwai2024}. 
The electromagnetic design and production were established, and frequency sweeping with two desired modes was successfully demonstrated. The overall performance of the measurement at 2.9~T was evaluated with Monte Carlo simulations.
In the future, a second dual-mode rectangular cavity for the measurement at 1.15~T field will also be developed to take advantage of the fact that the fluctuating magnetic field is minimal at this field because the derivative of $\nu_{12}^{\text{Mu}}$ and $\nu_{34}^{\text{Mu}}$ becomes zero.

For muonic helium HFS measurements at 1.15~T field, a cylindrical cavity with a single TM110 mode will be designed and constructed, similar to~\cite{Gardner1982}.
In the meantime, since the transition frequencies $\nu_{12}^{\mu\text{He}}$ and $\nu_{34}^{\mu\text{He}}$ at a field of 0.45~T and 2.9~T are nearly identical to those of muonium at 1.7~T, the same cylindrical microwave cavity will be used.

%Fig6---------------------------------------------------------
%-------------------------------------------------------------

%---------------------------------------
\subsection{Target chamber}

The target gas chamber containing the Kr gas (or He gas), common to both zero-field and high-field measurements, is made of pure aluminum with a 100~$\mu$m thick Al beam window 10 cm in diameter. 
The gas pressure can be changed from 0.3 to 1.0 atm and 1 to 2 atm.
For pressures above 1 atm, a small vacuum chamber with a 75~$\mu$m thick Kapton window mounted on the entrance window is used to avoid deforming the beam foil while evacuating the gas chamber. 
The gas pressure is measured by a pressure transducer (Fluke RPM4 reference pressure monitor) at 1~Pa level with an accuracy of 0.02\%.
The presence of contaminants is confirmed by quadrupole-mass spectrometry sampling the gas through a capillary tube before and after the measurement.
Oxygen is responsible for the depolarization of muonium through electron spin exchange collision. 
An oxygen contaminant of 4~ppm would correspond to an estimated signal drop of 7\%. 
Gas samples can also be collected in small cylinders during the measurement to monitor the gas purity. 
The climate control of the experimental area will keep the temperature variation below 0.1 degrees and be monitored during the experiment to reduce fluctuation in the gas pressure and microwave system.

For muonic helium HFS measurements, the entrance beam window of the gas chamber is made of a copper beryllium (CuBe) foil. The gas pressure can be changed from 1.0 to 10 atm by varying the CuBe foil thickness.
The He/CH$_{4}$ gas mixture is prepared by filling the first 2\% of the nominal pressure with CH$_{4}$ gas followed by high-purity He gas with an accuracy of 0.2\%. 
The relative He/CH$_{4}$ ratio between measurements and the presence of other contaminants is confirmed by quadrupole-mass spectrometry.

%---------------------------------------
\subsection{Detector system}

Several types of detectors are used in the measurements. 
A fiber beam profile monitor measures online the muon beam pulse by pulse, 
an offline three-dimensional profile monitor system is used to measure the muon beam stopping distribution in the Kr gas chamber, 
and an integrated highly-segmented scintillation detector system is used to detect positrons. Later, a new silicon strip detector was also tested and used.

%-------------------
\subsubsection{Beam profile monitor}

An online muon beam profile monitor is used to suppress systematic uncertainty related to the stability of the beam profile and intensity (Fig.~\ref{Fig7}(a)).
This profile monitor is made of $\phi$100~$\mu$m plastic scintillation fibers to be non-destructive as much as possible with a detection area of 100~mm $\times$ 100~mm to measure the muon beam pulse by pulse to reject irregular beam shape and pulse intensity. 
Each axis comprises 16 segments made of an array of 40 fibers bound together in a bundle (4~mm effective width), directly connected to a Silicon PhotoMultiplier (SiPM) at both ends, each segment spaced 2~mm apart and glued on a polyimide film 25~$\mu$m thick with epoxy resin for rigidity~\cite{Kanda2015,KandaThesis2017}. 
Fibers have less light reflection loss, and a better optical coupling to the SiPM can be achieved, resulting in smaller position dependence. 
The front-end electronics is composed of an EASIROC~\cite{Callier2012} chip that is used as an Amplifier Shaper Discriminator (ASD) and peak-hold Analog-to-Digital Converter (ADC). 
Muon beam profile and intensity measurements showed that higher efficiency with better uniformity along the fiber could be achieved as compared to a thin scintillator slab detector.

%-------------------
\subsubsection{Offline three-dimensional beam profile monitor}

The muon beam-stopping distribution in the gas chamber was measured offline by a three-dimensional beam monitor to suppress systematic uncertainty related to the muonium atom distribution in the chamber. 
This monitor is based on a muon beam profile monitor developed to diagnose pulsed muon beams at J-PARC MUSE~\cite{Ito2014}, which is composed of a scintillation screen, a gated image intensifier, and a cooled charge-coupled device (CCD) camera. 
The scintillation screen is placed inside the gas chamber (without the microwave cavity installed), which is closed with an acrylic plate at the back. 
The scintillation light from the screen is captured through the acrylic plate by a large aperture lens and sent to the image-intensified CCD unit. 
By moving the scintillation screen together with the detector unit, i.e., keeping the focal length constant, the muon-stopping distribution can be measured along the gas chamber. 
%The monitoring system is remotely controlled with an actuator coupled to a stepping motor to move the scintillation screen in the gas chamber, while the image intensifier and CCD camera mounted on a stage are moved simultaneously by a linear actuator. 
This system was used to measure muon beam distributions at several Kr gas densities before the zero-field experiments.
The precision of the beam center along the beam axis in the chamber was better than $\pm$~2~mm, corresponding to about 2~Hz uncertainty in the final HFS value based on a Monte Carlo simulation. 
%Details are reported in~\cite{Ueno2017a, Ueno2017b,UenoThesis2019}. 
Details are reported in~\cite{Ueno2017b,UenoThesis2019}. 
It will be used again to measure the beam-stopping distribution in the MRI magnet for the high-field experiments.

%-------------------
\subsubsection{Segmented scintillation detector }

Positrons (electrons) are detected with highly segmented scintillation counters to have high-rate capability due to the pulse nature of the muon beam. 
The total detection area is 240~mm $\times$ 240~mm, divided into 576 segments. Each segment consists of a small plastic scintillator (10~mm $\times$ 10~mm $\times$ 3~mm) with a SiPM directly attached.
The SiPMs are Multi-Pixel Photon Counters (MPPC) from Hamamatsu arranged in a 24 $\times$ 24 matrix on a printed circuit board. 
MPPC signals are directly processed by KALLIOPE (KEK Advanced Linear and Logic board Integrated Optical detector for Positron and Electron) fast readout electronics~\cite{Kojima2014}, which contains an Application Specific Integrated Circuit (ASIC) based ASD, and Field Programmable Gated Arrays (FPGA) based multi-hit Time-to-Digital Converter (TDC). 
The detector is composed of two layers, and a coincidence hit is required on two matching SiPMs to suppress accidental background due to dark noise. 
The maximum expected event rate is approximately 3 MHz, resulting in a pileup loss of 2\% of the total statistics from a simulation.
The pileup loss is mostly dominated by the analog pulse shape, which depends on the amplifier parameters. 
After pileup correction, related systematic uncertainty can be suppressed to an acceptable level. 
The uniform performance of the detector is ensured by tuning the operation voltage for each SiPM. 
This integrated detector system is described in detail in~\cite{Kanda2015, Kanda2016}.

%-------------------
\subsubsection{Silicon strip detector} 

\begin{figure}[t]
  \centering
  \includegraphics[width=76mm]{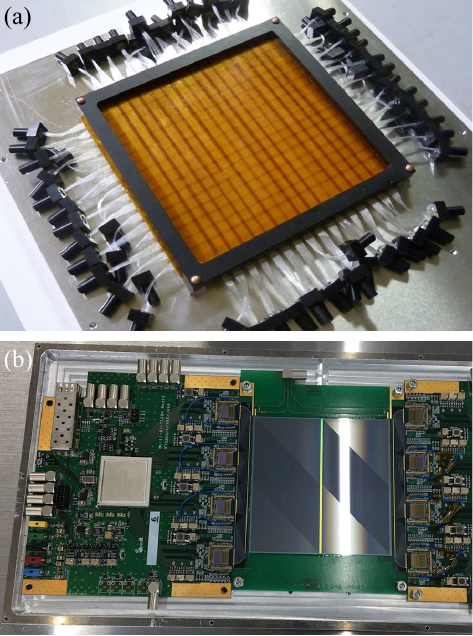}
  \caption{(a) Online muon beam profile monitor, and (b) integrated silicon strip detector. 
  %The sensor is directly connected on both sides to an analog/digital combined type integrated circuit for readout and signal processing through four ASIC- ASDs (Slit128A) controlled by a FPGA.
  }
  \label{Fig7}
\end{figure}

A new type of positron (electron) detector made of a highly-segmented silicon strip sensor with high-rate capability, which was originally developed for the J-PARC muon $g-2$/EDM experiment, was used and tested (Fig.~\ref{Fig7}(b))~\cite{NishimuraThesis2018}.
%Figure~\ref(Fig7}(b) shows a picture of this new integrated silicon strip detector. 
The silicon strip sensor has an active area of 97.28~mm square divided into two blocks with a thickness of 0.32~mm. The strip pitch and length are 0.19~mm and 48.575~mm, respectively. The number of strips is 512 $\times$ 2~blocks. 
The silicon strip sensor is connected on both sides through an adapter to two multi-Slit128A boards, where an analog/digital combined type integrated circuit directly processes signals. 
Each board contains four readout SliT128A chips, ASIC-based ASD, that are controlled by an FPGA. 
The details and performance of this silicon strip detector are reported in~\cite{Aoyagi2020}. 
This detector was used in an experiment at zero field in June 2018 to apply Rabi-oscillation spectroscopy to muonium HFS measurements (see Sects.~\ref{analysis} and \ref{results}, while for the high-field experiment, an improved version with four sensors mounted on one board and new readout ASIC chips SliT128D for better performance will be used.

%Fig7---------------------------------------------------------
%-------------------------------------------------------------

%---------------------------------------
\subsection{Blind analysis}

To ensure a data analysis free from any bias from previous experimental results, a blind analysis was selected based on the hidden answer method.
The true microwave frequency is hidden during the measurement by adding a \mbox{secret} offset value to the applied microwave frequency. To draw a resonance curve, this offset value must be fixed for all measurements at different frequencies, thus resulting in a measured resonance curve with that offset value. 
In our blind analysis, programs using passwords and encryption were developed.
A random offset value is first generated in the range $\pm$8~kHz, where the change in the stored microwave intensity is sufficiently small but still larger enough than the goal precision and encrypted with a password. 
The signal generator is operated using a separate program that requires a password at startup. 
The operating program decrypts the offset value and sets the frequency of the signal generator by adding the offset value to the input frequency. 
At the end of the measurements, the encrypted offset value data is deposited with someone outside of the collaboration. 
When the analysis is completed, the offset value data is received, and the blinds are opened using the decryption program and password. The blind program writer and the password administrator cannot perform the analysis.
Since two frequencies are to be measured, random offset values are generated separately for each transition frequency $\nu_{12}$ and $\nu_{34}$, which also allows a blind analysis for the muon-to-proton magnetic moment ratio obtained from the difference between those two frequencies.

\begin{figure}[t]
  \centering
  \includegraphics[width=76mm]{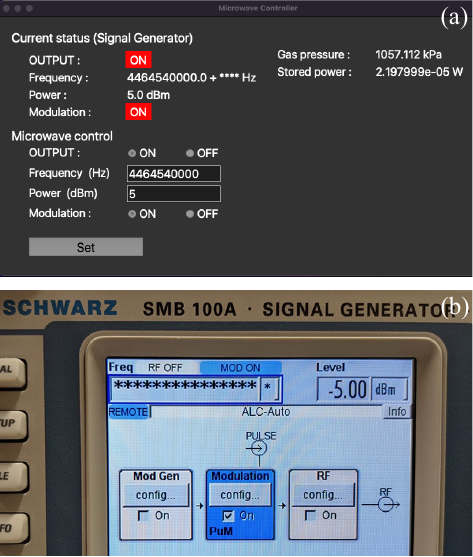}
  \caption{(a) Graphical User Interface of the signal generator operating program. (b) Display of the signal generator after the operation program started. The value of the applied microwave frequency is hidden (blinded).}
  \label{Fig8}
\end{figure}

To confirm the sequence of operations, this blind analysis program was applied to the HFS measurements of muonic helium atoms at zero field in February and May 2022~\cite{Strasser2023}. Figure~\ref{Fig8}(a) shows the Graphical User Interface of the signal generator operating program. This program also serves as a logger for the gas pressure and microwave power. After the program starts, the frequency display of the signal generator is hidden, as shown in Fig.~\ref{Fig8}(b).
When the analysis was completed, the blind was opened in May 2023, revealing a blind value of $-1,926$~Hz, thus completing the demonstration.

%Fig8---------------------------------------------------------
%-------------------------------------------------------------

%--------------------------------------------------------------------------------
\subsection{Analysis method}\label{analysis}

In standard spectroscopy, the resonance curve is drawn by sweeping the microwave frequency and measuring the spin-flip signal to determine the resonance frequency from the center of the resonance curve. 
In the predecessor experiment, a resonance line narrowing technique was adopted to improve the accuracy that only selects long-lived muonium atoms (``old muonium'' method)~\cite{Liu1999}. 
However, since the muon beam provided at LAMPF was quasi-continuous, 70\% of the muon flux was discarded to produce a pulsed time structure. Also, the ``old muonium'' technique was not applied in the analysis but rather implemented in the hardware.

At J-PARC, the muon beam is pulsed with a double-bunch structure 100~ns wide separated by 600~ns and repetitive at 25~Hz.
The measured positron data from $\mu^{+}$ decay (or electron data from $\mu^{-}$ decay for $\mu$He) will be recorded event-by-event and pulse-by-pulse for offline analysis. 
This will allow us to compare different methods to check the consistency of the analysis and investigate related systematic errors.
Pulsed muons will help reduce the background in the ``old muonium'' method. On the other hand, this analysis technique will also be an option to reduce the pileup if necessary. 

\begin{figure}[t]
  \centering
  \includegraphics[width=76mm]{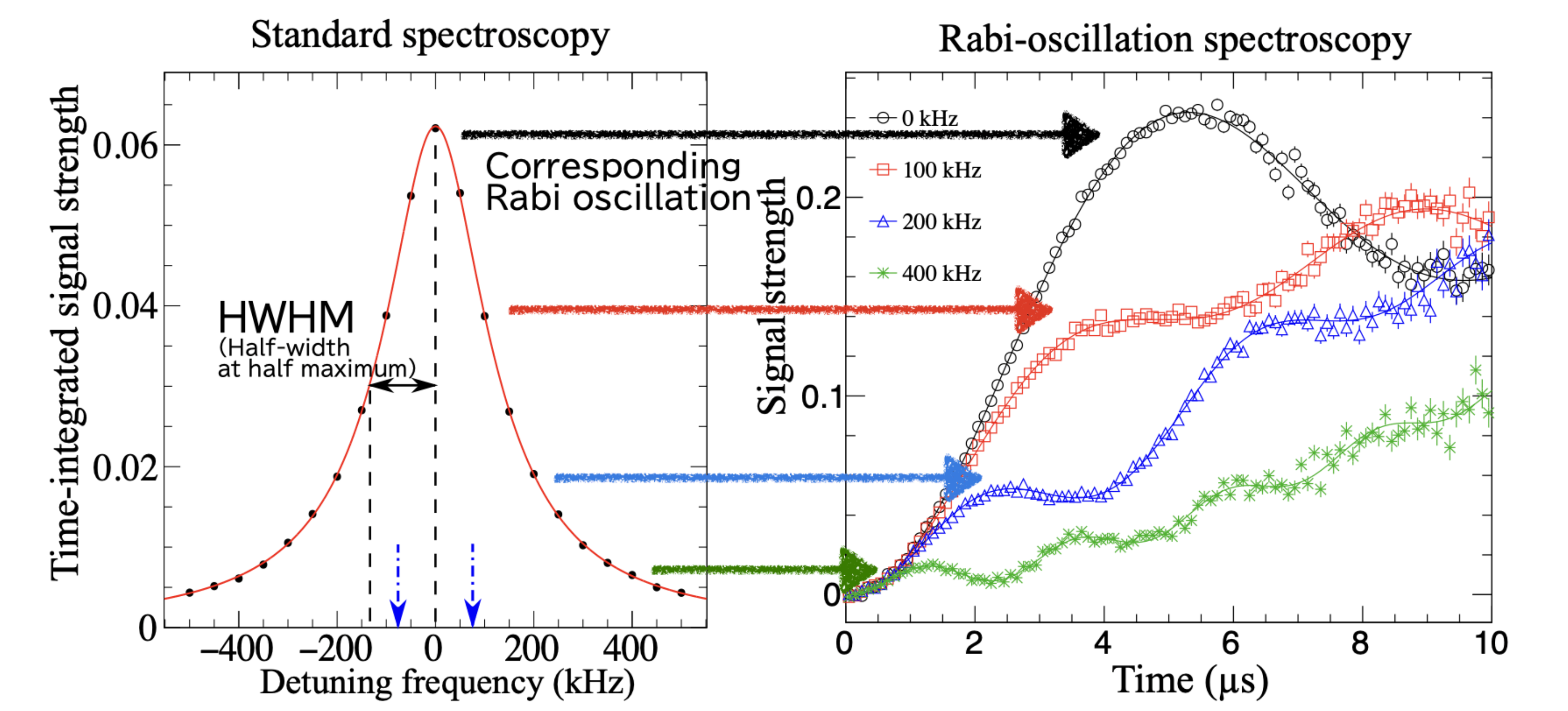}
  \caption{Resonance curve of the standard spectroscopy (left) and the time evolution of the Rabi oscillation used in the new spectroscopy obtained from a simulation (right).}
  \label{Fig9}
\end{figure}

However, it was pointed out that the Rabi-oscillation spectroscopy method can, in principle, extract more information than the standard time integral method but was unsuccessful in the previous measurements at LAMPF because of low statistics. 
In Rabi-oscillation spectroscopy, the resonance frequency is obtained from the time evolution of the Rabi oscillation obtained from a single measurement data without Fourier transform or microwave frequency sweep. 
The Rabi-oscillation spectroscopy method was developed and applied to muonium HFS measurements~\cite{NishimuraThesis2018, Nishimura2021}. 
A realistic 3D simulation study was performed to assert its potential by using the actual calculated microwave field distribution in the microwave cavity combined with the muon-stopping distribution (i.e., muonium distribution) in the Kr gas chamber. 
Figure~\ref{Fig9} shows a comparison between standard and Rabi-oscillation spectroscopy obtained from a simulation. 
Since the time evolution of the Rabi oscillation is uniquely determined by the microwave power and the resonance frequency, each information can be extracted simultaneously from only one detuning frequency data.
This method can determine resonance frequencies even when power fluctuates and thus is robust against systematic uncertainties due to power fluctuations (in the standard method, if the microwave intensity fluctuates during the sweep, the resonance curve becomes asymmetric, becoming a source of systematic uncertainty).
The most sensitive detuning frequency is found at $\sim$~60~kHz, and the statistical uncertainty can be improved by 3.2~times compared to the standard time integral method. 
This method requires high-statistics data for fitting and an accurate determination of the time and number of detected positrons (electrons). Thus, a detector with high-rate capability and good time resolution, like a silicon strip detector, is required.

%Fig9---------------------------------------------------------
%-------------------------------------------------------------

%--------------------------------------------------------------------------------
\section{Results at zero field}\label{results}

Several commissioning experiments at zero magnetic field have been performed since 2016 at the D2 area of MUSE \mbox{D-line}. 
The experimental apparatus shown in Fig.~\ref{Fig3} was enclosed in a magnetic shield box made of three permalloy layers to suppress the residual magnetic field. 
Several milestones were achieved.

%----------
%(1) Kanda2021
Initially, a small cylindrical microwave cavity with 81.8~mm inner diameter and 230~mm long was used with TM110 mode to match the muonium HFS transition frequency of 4.463 GHz~\cite{Tanaka2021}. 
The resonance was successfully observed at J-PARC for the first time using a high-intensity pulsed muon beam and a high-rate capable positron counter.
The muonium hyperfine structure interval $\Delta\nu_{\text{HFS}}^{\text{Mu}}$ = 4~463~302(4)~kHz (0.9~ppm) was obtained~\cite{Kanda2021}. 
%----------
The signal-over-noise ratio of that small cavity was poor due to a significant number of muons stopping in the wall. 
A larger and longer cylindrical cavity (181~mm inner diameter, 304~mm long) with TM220 mode was developed to enable muonium HFS measurement at lower gas pressure without severe statistics loss~\cite{UenoThesis2019}. This cavity was used in all subsequent experiments.

%----------
%(2) Nishimura2021
Rabi-oscillation spectroscopy was applied to silicon strip detector data obtained from an experiment in June 2018.
The resonance frequency was successfully determined from the Rabi oscillations obtained at multiple microwave frequencies. 
The result of the analysis gave $\Delta\nu_{\text{HFS}}^{\text{Mu}}$ = 4~463~301.61(71)~kHz (160~ppb)~\cite{Nishimura2021}, corresponding to nearly 2~times better precision from the previous zero-field experiment at LAMPF~\cite{Casperson1975}. 
This demonstrated that the target statistical precision could be achieved in the future at the H1 area, where the surface muon flux is estimated to be ten times higher than at the D2 area.

%----------
%(3) Strasser2023
Muonic helium HFS measurements were also performed using the same apparatus in 2021--2022.
Resonance curves were measured at three different He + CH$_{4}$(2\%) gas pressures of 3.0, 4.0, and 10.4~atm. 
Each measurement was performed under a user program in separate beam cycles.
The HFS frequency at zero pressure of a free $\mu$He atom was obtained by fitting a linear pressure shift to the data. 
The result of the analysis gave $\Delta\nu_{\text{HFS}}^{\text{Mu}}$ = 4~464~980(20)~kHz (4.5~ppm)~\cite{Strasser2023}.
After nearly 40 years, a new precise measurement with 3~times better precision than the previous measurement at zero field~\cite{Orth1980} was achieved. Our result is also more precise than the high-field measurement~\cite{Gardner1982}, improving the current world record by a factor of 1.5, and the first performed with CH$_{4}$ admixture to efficiently form neutral $\mu$He atoms.

%--------------------------------------------------------------------------------
\section{Experimental accuracy}

Considering the beam intensity of 1~$\times$~10$^8$~$\mu^+$/s at the J-PARC MUSE \mbox{H-line}, a statistical precision of 5~Hz (1.2~ppb) for $\Delta\nu_{\text{HFS}}^{\text{Mu}}$ is expected after 40 days of measurement.

%Tab1---------------------------------------------------------
\begin{table*}[h]
  \centering
\caption{Estimated systematic uncertainties in the high-field experiment.}
\label{Tab1}
%\begin{tabular}{@{}llll@{}}
\begin{tabular}{@{}lcccc@{}}
\toprule
\multicolumn{1}{l}{\textrm{Contributions}}&
\multicolumn{1}{l}{\textrm{Accuracy}}&
\multicolumn{1}{l}{\textrm{$\nu_{12}$, $\nu_{34}$}}&
\multicolumn{1}{l}{\textrm{$\delta$($\Delta\nu_{\text{HFS}}^{\text{Mu}}$)}}&
\multicolumn{1}{l}{\textrm{$\delta$($\mu_{\mu}/\mu_{p}$)}}\\
 & & (Hz) & (ppb) & (ppb)\\
\midrule
Magnetic field & 15~ppb &  & 0.0 & 8\\
Microwave power & $<$ 0.01\% &   $<$ 1 & 0.3 & 3\\
Kr gas temperature & 0.2~deg. & $<$ 2 & 0.4 &  4\\
Kr gas pressure & 1 Pa &   1 & 0.2 &  0\\
H impurity & $<$ 50~ppm &   1 & 0.5 &  0\\
Quadratic dependence (Mu) &   &   5 & 1.0 &  5\\
Muonium position (x,y) & 1~mm & 3 & 0.6 &  6\\
Muonium position (z) & 1~mm & $<$ 1 & 0.2 &  2\\
Beamline & 10$^{-4}$ & $<$ 1 & 0.2 &  2\\
Detector pile-up &  w/o absorber &  2.8 & 0.5 & 3\\
 & w/ absorber & 0.3 & $<$ 0.1 & $<$ 1\\
\midrule
Total &  &  & $\sim$ 1.5 & $\sim$ 13\\
\botrule
\end{tabular}
\end{table*}
%-------------------------------------------------------------

However, as we improve the statistics, the systematic uncertainty becomes more severe and needs to be carefully considered. 
A tool to estimate systematic uncertainty was developed~\cite{Tanaka2014, Torii2015} and is currently being used to investigate the effect of a particular fluctuation on the final result. 
The required precision is determined by a Monte Carlo simulation that calculates the resonance line from muonium distribution, microwave distribution, and detection efficiency.
The main sources of systematic uncertainty include, in order of importance, inhomogeneity in the magnetic field, microwave power, muon-stopping distribution, and gas density extrapolation. 
The current estimated systematic uncertainties are shown in Table \ref{Tab1}.

%----------
The magnetic field inhomogeneity and absolute measurement seem to be under control. 
According to the previous experiment, it has little effect on the determination of $\Delta\nu_{\text{HFS}}$ but is important in $\mu_{\mu}/\mu_{p}$. 
A measurement accuracy of $\sim$~15~ppb at 1.7~T would account for 8~ppb in $\delta$($\mu_{\mu}/\mu_{p}$). 

%----------
A fluctuation of 0.01 \% in the microwave power results in a shift of 1~Hz in the measured frequency, thus contributing 0.3~ppb to $\delta$($\Delta\nu_{\text{HFS}}$) and 3~ppb to $\delta$($\mu_{\mu}/\mu_{p}$), respectively.
A position drift of the muon beam distribution during the scan also affects the resolution, thus requiring constant monitoring of the beam profile.

%----------
The error from the beam profile monitor measuring every muon pulse with a precision of 1~mm would result in a shift of 3~Hz, or 0.6~ppb and 6~ppb, for $\delta$($\Delta\nu_{\text{HFS}}$) and $\delta$($\mu_{\mu}/\mu_{p}$), respectively. 
The effect on the longitudinal muon stopping distribution measured with equal precision by the offline three-dimensional muon beam monitor is less severe and would only account for a shift of 1~Hz, or 0.2~ppb and 2~ppb, respectively.

%----------
The quadratic dependence of the gas density extrapolation to the zero density limit will be improved by our ability to measure at lower Kr gas density with a longer cavity. 
The systematic uncertainty would be reduced to a shift of 5~Hz, thus contributing 1.0~ppb and 5~ppb, respectively. 

After considering all other error sources, the current error budget for systematic uncertainty in our experiment is at present estimated to be about $\sim$~1.5~ppb for $\delta$($\Delta\nu$) and $\sim$~13~ppb for  $\delta$($\mu_{\mu}/\mu_{p}$), respectively.

For muonic helium HFS measurements, most of the systematic uncertainties are common with muonium but are less severe due to the lower achievable precision. 
In addition, we need to consider the following contributions (see~\cite{Strasser2023} for details).
The uncertainty on the quadratic pressure dependence in helium is unknown because of not enough experimental data. 
An upper limit of 780~Hz was estimated. Additional measurements at high pressure are required to allow the determination of the quadratic pressure term. 
The uncertainty on the CH$_{4}$ concentration in helium is also difficult to ascertain because of the unknown value of its pressure shift. An upper value of $\sim$3~Hz/atm was estimated. However, this can be reduced by using the same mixture from a gas container for all measurements.

%--------------------------------------------------------------------------------
\section{Conclusion}\label{conclusion}

New precise muonium HFS measurements are of the utmost importance to test the QED theory and improve the determination of the positive muon magnetic moment and mass for new fundamental physics experiments. 
Complementary measurements of the muonic helium HFS will be beneficial to test and improve the theory of the three-body atomic system and to test further $CPT$ invariance by comparing the magnetic moments and masses of positive and negative muons. 
To achieve this goal, improving overall accuracy as well as estimating and understanding systematic uncertainties are crucial.

High-field measurements are now ready to collect data using the world's most intense pulsed muon beam at the MUSE \mbox{H-line}. 
In one day of measurements with muonium, we should already reach the same statistical precision as the previous experiment~\cite{Liu1999}.
Unfortunately, due to a problem with the neutron spallation target at the Materials and Life Science Experimental Facility (MLF) 
of J-PARC, where MUSE is located, scheduled beamtimes in June and December 2024 were unfortunately canceled.
We are now aiming at the first muonium HFS experiment at high field sometime in 2025.
Then, the first high-field measurements of muonic helium HFS will follow shortly.

%--------------------------------------------------------------------------------
%--------------------------------------------------------------------------------

\backmatter
%--------------------------------------------------------------------------------
%\bmhead{Supplementary information}

%If your article has accompanying supplementary file/s please state so here. 
%Please refer to Journal-level guidance for any specific requirements.

%--------------------------------------------------------------------------------
\bmhead{Acknowledgements}

The muon experiment at the Materials and Life Science Experimental Facility of J-PARC was performed under a user program (Proposals No. 2017A0134, 2017B0224, 2018A0071, 2018B0314, 2019A0144, 2020B004, 2020B0333, 2021B0169, 2022A0159).

%--------------------------------------------------------------------------------
\section*{Declarations}

%Some journals require declarations to be submitted in a standardised format. Please check the Instructions for Authors of the journal to which you are submitting to see if you need to complete this section. If yes, your manuscript must contain the following sections under the heading `Declarations':

\begin{itemize}

\item{\bf Funding Information}
This work was supported by the JSPS KAKENHI Grant No. 26247046, 17H01133, 19K14746, 21H04463, and 21H04481.

%Torii:     21H04463 Kiban-A
%Strasser:  21H04481 Kiban-A
%Nishimura: 19K14746 Wakate-kenkyu 
%Shimomura: 17H01133 Kiban-A
%Shimomura: 26247046 Kiban-A

%\item{\bf Competing Interests}
%The authors declare no competing interests.
%(not applicable)

%\item Ethics approval and consent to participate
%\item{\bf  Ethical Standards}
%(not applicable)

%\item{\bf  Consent for Publication}
%(not applicable)

%\item Data availability
\item{\bf  Data Availability Statement}
%There are no associated data with this article.
The data that support the findings of this study are available from the corresponding author upon reasonable request.

%\item{\bf Code Availability}
%(not applicable)

\item{\bf Author Contributions}
%(not applicable)
All authors contributed equally to this work.
%(sample) All authors contributed to the studyÕs conception and design. Material preparation, data collection and analysis were performed by xx, yy and zz. The first draft of the manuscript was written by xx. All authors commented on previous versions of the manuscript. All authors read and approved the final manuscript.

\end{itemize}

\bibliography{MuSEUM_biblio}% common bib file

\end{document}